\documentclass[letter,twocolumn]{jpsj2}

\usepackage{times}
\usepackage{graphicx}

\title{Inverse Isotope Effect on Kondo Temperature
in Electron-Rattling System}

\author{Takashi {\sc Hotta}}

\inst{Department of Physics, Tokyo Metropolitan University,
1-1 Minami-Osawa, Hachioji, Tokyo 192-0397}

\recdate{\today}

\abst{
In an electron system coupled with anharmonic phonons,
i.e., {\it rattling},
inverse isotope effect on the Kondo temperature $T_{\rm K}$ is
found to occur by the numerical evaluation of
the Sommerfeld constant $\gamma$ of the Anderson-Holstein model.
For the anharmonic potential of an oscillator with mass $M$
in which large $\gamma$ has been found to be almost independent
of an applied magnetic field,
$\gamma$ is significantly suppressed when $M$ is increased,
i.e., $T_{\rm K}$ is enhanced due to the relation of
$\gamma$$\sim$$T_{\rm K}^{-1}$ in the Kondo problem,
leading to the inverse isotope effect on $T_{\rm K}$.
Since this phenomenon does not occur for harmonic phonons,
it can be a key experiment to prove the relevance of
rattling to magnetically robust heavy electron state.
}

\kword{Rattling, Kondo effect, Isotope effect}

\begin{document}
\maketitle


In traditional heavy-fermion materials such as CeCu$_6$,
\cite{Stewart}
the electronic specific heat coefficient $\gamma$
is significantly suppressed by an applied magnetic field.
This can be understood within the standard Kondo effect,
\cite{Kondo1}
i.e., the singlet formation of local magnetic moment
due to the coupling with conduction electrons.\cite{Yosida}
From the viewpoint of the standard Kondo effect
originating from spin degree of freedom,
it is quite unusual that heavy effective mass is independent
of the applied magnetic field.
Thus, the magnetically robust heavy-fermion phenomenon
observed in SmOs$_4$Sb$_{12}$~\cite{Sanada}
has attracted much attention of condensed-matter theorists
due to the renewed interest on the realization of
non-magnetic Kondo effect.

The theoretical research of the Kondo effect has
a history longer than forty years.\cite{Kondo40}
It has been widely recognized that the Kondo-like phenomenon
generally occurs in a conduction electron system hybridized with
a localized entity with internal degrees of freedom.
For instance, when electrons are coupled with Einstein phonons,
a double-well potential is formed in an adiabatic approximation,
leading naturally to a two-level system,
in which Kondo has first considered a possibility of
non-magnetic Kondo behavior.\cite{Kondo2a,Kondo2b}
This two-level Kondo system has been shown to exhibit
the same behavior as the magnetic Kondo effect.\cite{Vladar1,Vladar2}
In the context of heavy-fermion behavior in A15 compounds,
Kondo effect with phonon origin has been discussed.
\cite{Miyake1a,Miyake1b}
Further efforts to include the low-lying levels of local phonon
has been done to understand the quasi-Kondo behavior
in electron-phonon systems.\cite{Miyake2}

Due to the discovery of the magnetically robust
heavy-fermion phenomenon in SmOs$_4$Sb$_{12}$,
the research on the Kondo effect with phonon origin
seems to revive in recent years.
This phenomenon occurs on the filled skutterudite structure,
in which rare-earth guest atom is
surrounded by the cage composed of twelve pnictogens.
Then, the rare-earth ion easily moves around potential minima
or wide bottom of the potential inside the pnictogen cage.
This anharmonic motion is called rattling,
which is believed to have significant effect on
electronic properties of filled skutterudites.

As an extension of the two-level Kondo problem,
the four- and six-level Kondo systems have been analyzed
to clarify the effect of rattling in filled skutterudite.
\cite{Hattori1,Hattori2}
The origin of heavy mass has been discussed on the basis of the
Anderson-Holstein model.\cite{Hewson,Jeon,Lee,Ono,Hotta1}
The present author has discussed the magnetic field dependence of
Sommerfeld constant in the Anderson-Holstein model.\cite{Hotta2}
Then, it has been shown that large $\gamma$ becomes magnetically
robust when the bottom of the potential becomes flat due to
the effect of anharmonicity.
From these efforts, it has been gradually understood that
magnetically robust heavy-fermion phenomenon actually occurs
in electron-rattling systems, but we do not have an evidence
for the relevance of rattling to this intriguing phenomenon.
Thus, it is meaningful to consider an experiment
which confirms the contribution of rattling to the Kondo effect.

In this letter, we show that inverse isotope effect
occurs in the magnetically robust heavy electron state.
For the purpose, we numerically evaluate the Sommerfeld
constant $\gamma$ of the Anderson model
coupled with local oscillation of guest atom.
For the anharmonic potential in which $\gamma$ is almost independent
of an applied magnetic field,
when we increase the mass of the oscillator $M$,
we observe the decrease of $\gamma$,
i.e., the increase of the Kondo temperature $T_{\rm K}$
due to the relation of $\gamma$$\sim$$T_{\rm K}^{-1}$
in the Kondo problem.
Namely, in the expression of $T_{\rm K}$$\propto$$M^{-\eta}$,
the exponent $\eta$ becomes negative.
This is the inverse isotope effect on $T_{\rm K}$,
which can be an evidence for the relevance of anharmonic phonons
to the Kondo effect,
since it does not occur for harmonic phonons.


Let us first discuss how the isotope effect appears in
the local electron-rattling state.
The local term is expressed as
$H_{\rm loc}$=$U n_{\uparrow}n_{\downarrow}+g n Q+P^2/(2M)+V(Q)$,
where $U$ denotes Coulomb interaction,
$n_{\sigma}$=$d^{\dag}_{\sigma}d_{\sigma}$,
$d_{\sigma}$ is an annihilation operator of localized electron
with spin $\sigma$ at an impurity site,
$g$ is the electron-phonon coupling constant,
$n$=$n_{\uparrow}$+$n_{\downarrow}$,
$Q$ is normal coordinate of the oscillator,
$P$ is the corresponding canonical momentum,
$M$ is the mass of the oscillator,
and $V(Q)$ is the potential for the oscillator.
Here we express $V(Q)$ as
$V(Q)$=$k Q^2/2+k_4Q^4+k_6Q^6$,
where $k$ is a spring constant and we consider fourth- and
sixth-order anharmonicity, $k_4$ and $k_6$, in the potential.
Note that we use such units as $\hbar$=$k_{\rm B}$=1 in this paper.

By following the standard procedure of quantization of phonons,
we introduce the phonon operator $a$ defined
through $Q$=$(a+a^{\dag})/\sqrt{2\omega M}$,
where $\omega$ is the phonon energy, given by $\omega$=$\sqrt{k/M}$.
Then, we obtain
\begin{equation}
  \begin{split}
   H_{\rm loc} &=U n_{\uparrow}n_{\downarrow}
   + \omega \sqrt{\alpha}(a+a^{\dag}) n + \omega(a^{\dag}a+1/2) \\
   &+ \beta_4 \omega (a+a^{\dag})^4+\beta_6 \omega (a+a^{\dag})^6,
  \end{split}
\end{equation}
where $\alpha$ indicates the non-dimensional electron-phonon
coupling constant defined by $\alpha$=$g^2/(2M\omega^3)$
and non-dimensional anharmonicity parameters,
$\beta_4$ and $\beta_6$,
are given by $\beta_4$=$k_4/(4M^2\omega^3)$ and
$\beta_6$=$k_6/(8M^3\omega^4)$, respectively.

Here we explain $M$ dependence of parameters.
First we note the relation $\omega \propto M^{-1/2}$,
since we usually consider that the spring constant $k$
does not depend on $M$.
From the definition, $\alpha$ itself depends on $M$ as
$\alpha \propto M^{1/2}$.
Then, $\alpha \omega$ does not depend on $M$.
Here it is instructive to recall the isotope effect
in BCS superconductor.
Namely, the $M$ dependence of superconducting transition
temperature $T_{\rm c}$ occurs only through $\omega$,
since the effective attraction between electrons mediated
by harmonic phonons is given by $\alpha \omega$.
Then, we obtain the famous formula
$T_{\rm c} \propto M^{-1/2}$
for the BCS superconductor.
Concerning anharmonicity parameters,
we also consider that $k_4$ and $k_6$ are independent of $M$.
Thus, we obtain that $\beta_4 \propto M^{-1/2}$ and
$\beta_6 \propto M^{-1}$.

In this paper, we set
${\tilde \omega}$=$\omega/\sqrt{m}$,
${\tilde \alpha}$=$\alpha\sqrt{m}$,
${\tilde \beta}_4$=$\beta_{4}/\sqrt{m}$,
and ${\tilde \beta}_6$=$\beta_{6}/m$,
where $m$ denotes the mass ratio of the oscillator
when we substitute the oscillator atom with its isotope.
The quantities without tilde denote those for $m$=1.
For the actual calculations, we define the phonon basis as
$|\ell \rangle$=$(a^{\dag})^{\ell}|0\rangle/\sqrt{\ell!}$,
where $\ell$ is the phonon number and
$|0\rangle$ is the vacuum state.
The phonon basis is truncated at a finite number,
which is set as 1000 in this paper.

We briefly explain the change of the potential shape
due to $\beta_{4}$.\cite{Hotta2}
The potential is rewritten as
$V(q)$=
$\alpha\omega(q^2+16\alpha \beta_4 q^4+64 \alpha^2 \beta_6 q^6)$,
where $q$ is the non-dimensional length given by $q$=$Q\omega^2/g$.
Note that $V(q)$ is independent of $M$,
since ${\tilde \alpha}{\tilde \omega}$=$\alpha\omega$,
${\tilde \alpha} {\tilde \beta}_4$=$\alpha \beta_{4}$, and
${\tilde \alpha}^2 {\tilde \beta}_6$=$\alpha^2 \beta_{6}$.
In Fig.~1(a), we show $V(q)/(\alpha\omega)$ for several values of
$\beta_{4}$ for $\alpha$=2 and $\beta_{6}$=$10^{-5}$.
For $\beta_{4}$=0, $V(q)$ has a single minimum at $q$=0.
When we decrease $\beta_{4}$,
shoulder-like structure begins to appear around $q$=$\pm 4$
and the bottom of the potential becomes relatively wide.
For $\beta_{4}$$<$$-0.00274$, potential minima at $q$$\ne$0 appear.
When $\beta_{4}$ is smaller than $-0.003$,
a couple of minima at $q$$\ne$0 are gradually deep.
In the following, the potential
in the region of $-0.003$$<$$\beta_4$$<$$-0.002$
is called {\it the rattling type}.

As mentioned above, there occurs attractive interaction
$U_{\rm ph}$ between electrons mediated by phonons.
The effective interaction $U_{\rm eff}$ is evaluated by
$U_{\rm eff}$=$U-U_{\rm ph}$=$E^{(0)}_0+E^{(0)}_2-2E^{(0)}_1$,
where $E^{(0)}_{n}$ is the ground-state energy of
$H_{\rm loc}$ for local electron number $n$.
For the case of harmonic potential
($\beta_{4}$=$\beta_{6}$=0),
$U_{\rm ph}$ is analytically evaluated as
$U_{\rm ph}$=$2\alpha\omega$.
Thus, we obtain $U_{\rm eff}$=$U-2\alpha\omega$,
which does not depend on $M$,
when we simply assume that $U$ is not affected by $M$.

On the other hand, for the anharmonic potential,
significant $M$ dependence appears in $U_{\rm eff}$,
which will be a source of the isotope effect on
the Kondo temperature.
In order to visualize the $M$ dependence of $U_{\rm eff}$,
we consider the variation of $U_{\rm eff}$ due to the increase of $M$,
given by $\delta U_{\rm eff}$=$U_{\rm eff}(m)-U_{\rm eff}(1)$,
where $U_{\rm eff}(m)$ is the effective interaction for $m$.
Since $U$ is assumed to be unchanged by $M$,
the variation occurs only through the change of $U_{\rm ph}$.

In Fig.~1(b), we plot $\delta U_{\rm eff}$ as a function of
$\beta_{4}$ for $m$=1.05, $\beta_{6}$=$10^{-5}$, $\alpha$=2,
and $U/\omega$=10.
For harmonic potential, $\delta U_{\rm eff}$ is zero,
as shown by the red line.
For anharmonic potential, $U_{\rm eff}$ is decreased,
i.e., $U_{\rm ph}$ is increased, due to the increase of $M$
for the potential with single minimum,
while $U_{\rm eff}$ is increased with the increase
of $M$ for the potential with a couple of
deep minima at $q \ne 0$.
For the rattling-type potential,
the variation of $U_{\rm ph}$ due to the increase of $M$
is relatively large, since the flat potential easily
deforms by electron-phonon coupling.

\begin{figure}[t]
\label{fig1}
\centering
\includegraphics[width=8.5truecm]{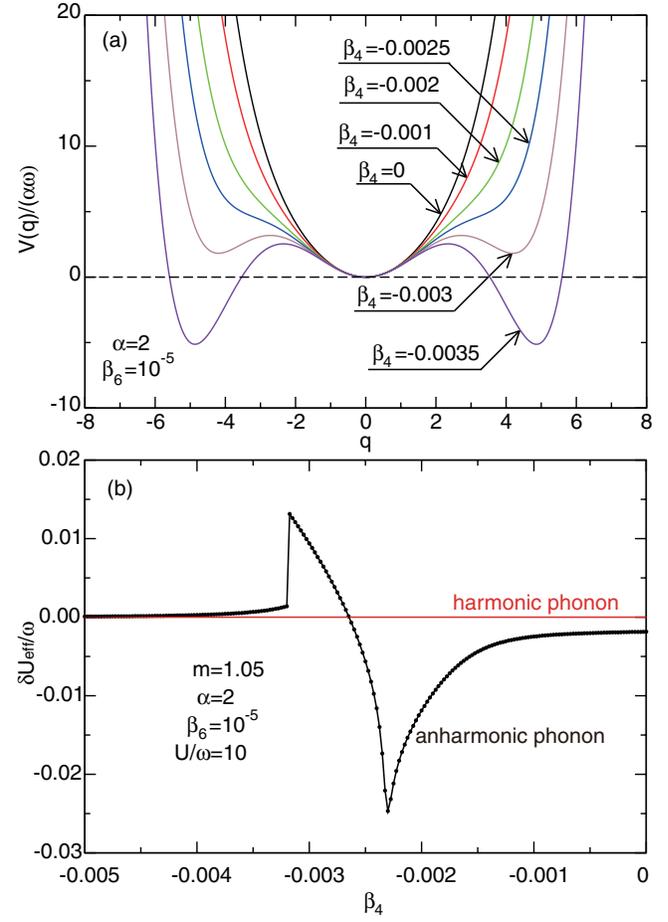}
\caption{(Color online)
(a) Anharmonic potential vs. $q$
for several values of $\beta_{4}$.
Other parameters are $\beta_{6}$=$10^{-5}$ and $\alpha$=2.
(b) Variation of the effective interaction
$\delta U_{\rm eff}$ as a function of $\beta_{4}$
for $m$=1.05, $\beta_{6}$=$10^{-5}$, $\alpha$=2, and
$U/\omega$=10.
The red line denotes the result for harmonic potential.
}
\end{figure}


Let us now consider the hybridization between localized and
conduction electrons.
The Hamiltonian is called the Anderson-Holstein model,
given by
\begin{equation}
   H \!=\! \sum_{\mib{k}\sigma} \varepsilon_{\mib{k}}
   c_{\mib{k}\sigma}^{\dag} c_{\mib{k}\sigma}
   \!+\! \sum_{\mib{k}\sigma}
   (Vc_{\mib{k}\sigma}^{\dag}d_{\sigma}+{\rm h.c.})
   \!+\! \mu n \!+\! H_{\rm loc},
\end{equation}
where $\varepsilon_{\mib{k}}$ denotes the dispersion of conduction electron,
$c_{\mib{k}\sigma}$ is an annihilation operator of conduction electron
with momentum $\mib{k}$ and spin $\sigma$,
$\mu$ is a chemical potential,
and $V$ is the hybridization between conduction and localized electrons.
In the following, we set $V/D$=0.25,
where $D$ is a half of the conduction bandwidth.
Note that we adjust $\mu$ to consider the half-filling case,
even though we do not mention explicitly.

For the evaluation of the Sommerfeld constant $\gamma$
of the Anderson-Holstein model, 
we employ a numerical renormalization group (NRG) method,
\cite{NRG}
where we include efficiently the conduction
electron states near the Fermi energy by discretizing
momentum space logarithmically.
In the actual calculation, we introduce a cut-off $\Lambda$
for the logarithmic discretization of the conduction band.
Due to the limitation of computer resources,
we usually keep only $L$ low-energy states.
In this paper, we set $\Lambda$=5 and $L$=2000.
By using the NRG method,
we calculate the Sommerfeld constant $\gamma$
at a low temperature $T$.
Note that $T$ is defined as $T_N$=$\Lambda^{-(N-1)/2}$
in the NRG calculation, where $N$ denotes the number of
the renormalization step.
We actually evaluate $\gamma$
through the relation of $\gamma$=$C/T$ at a low temperature $T$,
where $C$ is the specific heat of localized electron.
Namely, it is necessary to obtain $C$ with high accuracy.
For the purpose, we evaluate it by the numerical derivative of
the entropy $S$ through the relation of
$C$=$\partial S /\partial \log T$=$
(S_{N-1}-S_N)/\log (\Lambda^{1/2})$,
where $S_N$ is the entropy at the step $N$.

Here we summarize the previous NRG results of
$\gamma$ for $m$=1.\cite{Hotta2}
We have found that phonon-mediated attraction is largely
enhanced for $-0.0027$$<$$\beta_{4}$$<$$-0.0021$,
in which the potential has wide bottom and $\delta U_{\rm eff}$
is negative with large absolute value.
There occurs a strong cancellation between Coulomb repulsion and
the phonon-mediated attraction,
just when the potential shape is deformed from the rattling type.
We have shown that in such a situation,
spin and charge fluctuations are comparable to each other,
leading to the emergence of electron-rattling complex state
with large and magnetically robust $\gamma$.
In the present notation, magnetically robust large $\gamma$
has been found around at $\beta_{4}$$\approx$$-0.0021$ and
$-0.0027$.

\begin{figure}[t]
\label{fig2}
\centering
\includegraphics[width=8.5truecm]{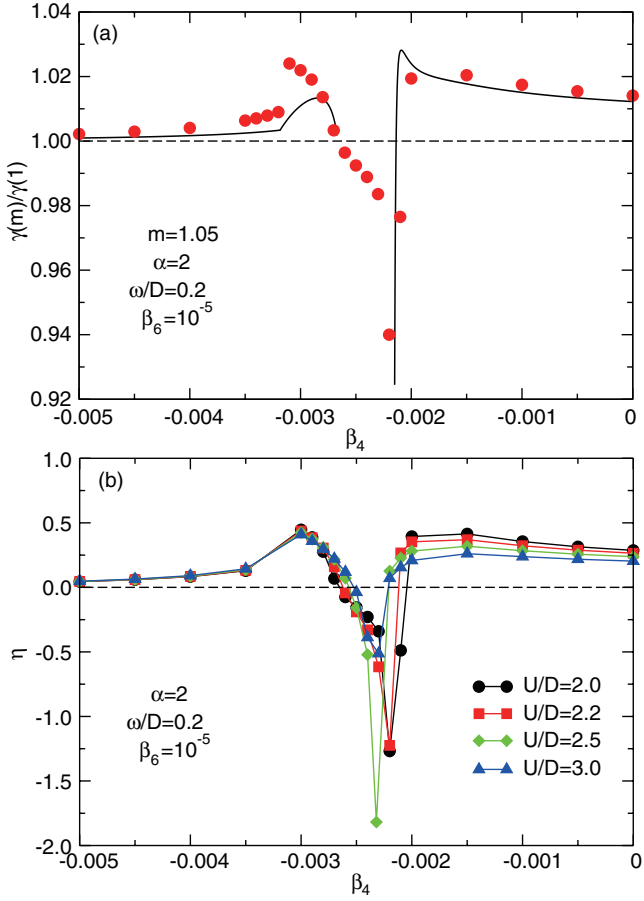}
\caption{(Color online)
(a) Ratio of the Sommerfeld constant $\gamma(m)/\gamma(1)$
vs. $\beta_{4}$ for $U/D$=2, $\alpha$=2, $\omega/D$=0.2,
$\beta_{6}$=$10^{-5}$, and $m$=1.05.
Numerical results are shown by red circles.
Solid curve is obtained from the analytic expression
of $T_{\rm K}$ of the effective $s$-$d$ model,
except for the region with $U_{\rm eff}<0$.
(b) Exponent $\eta$ in the relation of $T_{\rm K} \propto M^{-\eta}$
vs. $\beta_{4}$ for several values of $U$.
Other parameters are $\alpha$=2, $\omega/D$=0.2,
and $\beta_{6}$=$10^{-5}$.
}
\end{figure}

Now we consider the isotope effect on $\gamma$
for rattling phonons.
In Fig.~2(a), we show $\beta_{4}$ dependence of
$\gamma(m)/\gamma(1)$ for $U/D$=2, $\alpha$=2, $\omega/D$=0.2,
and $m$=1.05,
where $\gamma(m)$ is the Sommerfeld constant for $m$.
For $\beta_{4}$$>$$-0.0021$ and $\beta_{4}$$<$$-0.0027$,
we observe $\gamma(m)/\gamma(1)$$>$1.
For $-0.0021$$>$$\beta_{4}$$>$$-0.0027$
with relatively wide bottom in the potential,
when $m$ is increased, $\gamma$ is decreased,
i.e., $T_{\rm K}$ is increased due to the relation of
$\gamma$$\sim$$T_{\rm K}^{-1}$ in the Kondo problem.\cite{review}
This is the inverse isotope effect,
since in the standard isotope effect,
the relevant temperature is decreased with the increase of $M$,
as observed in the BCS superconductor.

The inverse isotope effect
is caused by the further decrease of negative effective interaction
$U_{\rm eff}$ in the region of $-0.0021$$>$$\beta_{4}$$>$$-0.0027$
due to the increase of $M$, as found in Fig.~1(b).
Note that $\beta_{4}$$>$$-0.0021$,
$\delta U_{\rm eff}$ is negative,
but $U_{\rm eff}$ is positive with large absolute value.
In this case, the change in $T_{\rm K}$ is determined
by the exchange interaction, not by $U_{\rm eff}$.
When $U_{\rm eff}$ is positive and is larger than
the width of the virtual bound state,
the Anderson-Holstein model is reduced to
the isotropic $s$-$d$ model with the exchange
interaction $J$,\cite{Hotta1} expressed as
\begin{equation}
 J = V^2 \sum_{\ell=0}^{\infty}
  \Biggl[
  \frac{|\langle \Phi_0^{(\ell)} | \Phi_1^{(0)} \rangle|^2}
  {E^{(\ell)}_0-E^{(0)}_1-\mu} + 
  \frac{|\langle \Phi_2^{(\ell)} | \Phi_1^{(0)} \rangle|^2}
  {E^{(\ell)}_2+\mu-E^{(0)}_1}
  \Biggr],
\end{equation}
where $|\Phi_n^{(\ell)} \rangle$ is the $\ell$-th eigenstate
of $H_{\rm loc}$ for electron number $n$
with the eigenenergy $E_n^{(\ell)}$.
Then, $T_{\rm K}$ is given by the well-known formula
$T_{\rm K}=D e^{-1/(2\rho J)}$,
where $\rho$ is the density of states at the Fermi energy.
The solid curve in Fig.~2(a) indicates
$[T_{\rm K}(m)/T_{\rm K}(1)]^{-1}$.
Except for the region with negative $U_{\rm eff}$,
the numerical results are reproduced
by the rough estimation of $T_{\rm K}$.
For $U_{\rm eff}$$<$0, analytic expression
of $T_{\rm K}$ is not obtained,
but we observe the tendency in the solid curve
toward the decrease of $[T_{\rm K}(m)/T_{\rm K}(1)]^{-1}$.
Thus, it is believed that the inverse isotope effect
is the essential feature of the anharmonic potential.

In Fig.~2(b), we show the $\beta_{4}$ dependence
of the exponent $\eta$ in the formula of the isotope effect
$T_{\rm K} \propto M^{-\eta}$.
For the evaluation of $\eta$,
it is convenient to use the relation
$\eta$=$\log [\gamma(m)/\gamma(1)]/\log m$
for $m$ near unity.
Here we set $m$=1.05.
When $\gamma(m)/\gamma(1)$ is smaller than unity,
the exponent becomes negative,
directly suggesting the inverse isotope effect.
When we increase $U$, the region with negative $\eta$ is
not so rapidly suppressed, as found in the results
for $U/D$=2.2 and 2.5.
In particular, for $U/D$=3, $U_{\rm eff}$ is found to be
always positive
in the present parameters, but we still find the negative $\eta$.
Thus, the effective attraction is not the only condition
for the appearance of the inverse isotope effect.
Rather, the key issue is the heavy electron state dressed by
rattling phonons,
when the potential bottom becomes wide.
From the above results, $\eta$ is negative only
for the rattling-type potential, including the region
with magnetically robust $\gamma$.


In order to confirm that the inverse isotope effect
occurs only for anharmonic phonons,
we discuss the results for harmonic phonons,
i.e., $\beta_{4}$=$\beta_{6}$=0.
Roughly speaking, the electronic state is understood
by the comparison of $U$ and $2\alpha\omega$.
For $U > 2\alpha\omega$, the local electron state
has double degeneracy of spin degree of freedom, even though
the Coulomb interaction is weakened by the attraction
mediated by phonons.
In this region, the model is described by the $s$-$d$
model with the exchange interaction $J$.
On the other hand, for $U < 2\alpha\omega$,
the vacant and double-occupied states are degenerate,
leading to the situation of the two-level Kondo system.
In such a case, the Anderson-Holstein model
is effectively reduced to the anisotropic $s$-$d$ model,
in which the longitudinal exchange interaction $J_1$
is different from the transverse component $J_2$.
Due to the immobility of bi-polaron in comparison
with single polaron, $J_2$ is smaller than $J_1$.
As for more details,
readers can consult with Ref.~\citen{Hotta1}.

\begin{figure}[t]
\label{fig3}
\centering
\includegraphics[width=8.5truecm]{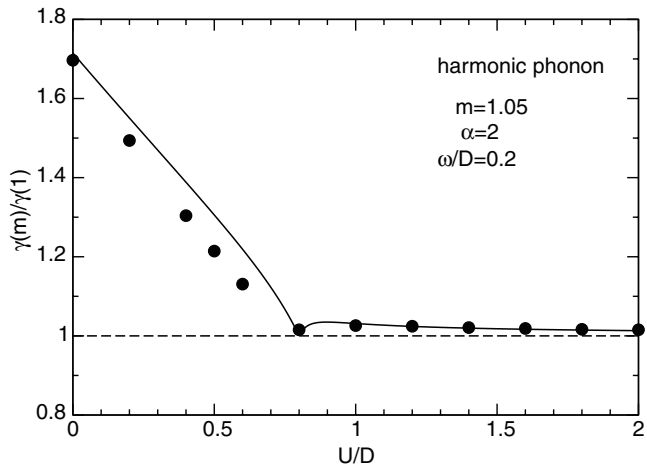}
\caption{$\gamma(m)/\gamma(1)$ vs. $U$
for $\beta_{4}$=$\beta_{6}$=0,
$m$=1.05, $\alpha$=2, and $\omega/D$=0.2.
Solid curve is estimated from $T_{\rm K}$
of the effective $s$-$d$ model.\cite{Hotta1}
}
\end{figure}

In Fig.~3, we depict the $U$ dependence of
$\gamma(m)/\gamma(1)$
for $m$=1.05, $\alpha$=2, $\omega/D$=0.2,
and $\beta_4$=$\beta_6$=0.
For comparison, we depict the curve obtained from
the analytic forms of $T_{\rm K}$ of the effective
$s$-$d$ models.\cite{Hotta1}
Since the expressions of $T_{\rm K}$ are obtained in
the strong-coupling limit,
the numerical results should not perfectly agree with
the analytic ones.
Nevertheless, the overall behavior is well reproduced by
the solid curve,
even though the present situation is not
in the strong-coupling limit.
It is not surprising that
the isotope effect is also found in
$T_{\rm K}$ for harmonic phonons,
since the virtual phonon excitations are included in the
expressions for $J$, $J_1$, and $J_2$.
Here we emphasize that $\gamma(m)/\gamma(1)$ is always
increased, namely, $T_{\rm K}(m)/T_{\rm K}(1)$ is
decreased, when we increase $m$.
Thus, the isotope effect on $T_{\rm K}$ appears
for harmonic phonons, but $T_{\rm K}$ is always decreased
with the increase of $m$.


In order to detect the inverse isotope effect,
we should estimate experimentally
$\gamma$ from the specific heat measurement,
e.g., of SmOs$_4$Sb$_{12}$ with isotope of Sm,
since we consider the oscillation of guest atom.
The idea is simple, but it is challenging
to find the direct evidence of the relevance of
rattling to the Kondo effect.
In the present paper, we have considered the isotope effect
on $\gamma$, but we have also checked that
the same effect appears
in magnetic susceptibility $\chi$, although
the results are not shown here.
Since the correspondence between $\chi$ and $T_{\rm K}$ is
clear in general, it may be more practical to discuss
experimentally the isotope effect on $\chi$.
In any case, we believe that the change of $\gamma$
and $\chi$ in the compound including isotope is
detectable in experiments.


Three comments are in order.
(i) We should pay due attention to the effect of local interaction,
when we discuss the actual
periodic system from the impurity model.
For instance, the magnitude of $\eta$ in the periodic system
may not be so large in comparison with the present results.
The point is that $\eta$ becomes negative
for magnetically robust heavy electron state.
(ii) When we substitute Sm with its isotope in SmOs$_4$Sb$_{12}$,
$T_{\rm K}$ may be distributed,
which makes it difficult to detect the isotope effect
on $\gamma$ and $\chi$.
This point seems to be related to the improvement of sample quality
due to the increase of filling fraction of rare earth atom in the
pnictogen cage of filled skutterudites.\cite{Tanaka}
(iii) In general, $T_{\rm K}$ should be distinguished from
the so-called coherence temperature concerning the formation
of heavy electron state.
The relation between $T_{\rm K}$ and observables
is not so clear in actual heavy-electron materials.
When rattling is relevant to the heavy electron state,
the isotope effect may significantly appear
in the coherence temperature.
This point may be discussed on the basis of
the periodic Anderson-Holstein model,
but it is one of future issues.


In summary, the Sommerfeld constant $\gamma$
of the Anderson-Holstein model
has been evaluated by the NRG method.
For the rattling-type potential of the oscillator,
when we increase the mass of the oscillator,
we have found the decrease of $\gamma$,
i.e., the increase of $T_{\rm K}$.
The same effect also occurs in the magnetic susceptibility $\chi$.
Then, we have concluded that the inverse isotope effect
on $T_{\rm K}$ occurs, if the magnetically robust large $\gamma$
originates from rattling phonons.
We expect that the measurements of $\gamma$ and $\chi$
in SmOs$_4$Sb$_{12}$ including Sm isotope will be performed in future.


The author thanks Y. Aoki for discussions.
This work has been supported by a Grant-in-Aid for
for Scientific Research on Innovative Areas ``Heavy Electrons''
(No. 20102008) of The Ministry of Education, Culture, Sports,
Science, and Technology, Japan.
The computation in this work has been done using the facilities
of the Supercomputer Center of Institute for Solid State Physics,
University of Tokyo.


\end{document}